\documentclass[final,1p,times]{elsarticle}

 \usepackage{graphics}
 \usepackage{epsfig}

\usepackage{amssymb}

\def\be {\begin{equation}}
\def\ee {\end{equation}}
\def\bea {\begin{eqnarray}}
\def\eea {\end{eqnarray}}
\def\nn {\nonumber}

\journal{Physics Letters B}

\begin{document}

\begin{frontmatter}

\title{Energy loss due to field fluctuations in a two-stream
QCD plasma}

 \author[label1]{Mahatsab Mandal}
 \address[label1]{mahatsab.mandal@saha.ac.in}
\author[label2]{Sreemoyee Sarkar}
 \address[label2]{sreemoyee.sarkar@saha.ac.in}
\author[label3]{Pradip Roy}
 \address[label3]{pradipk.roy@saha.ac.in}
\author[label4]{Abhee K. Dutt-Mazumder}
 \address[label4]{abhee.dm@saha.ac.in}
\address{High Energy Nuclear and Particle Physics Division, Saha Institute of Nuclear Physics,
 1/AF Bidhannagar, Kolkata-700 064, INDIA}

\begin{abstract}
We derive the expression for the collisional energy loss in two stream plasma induced by the 
fluctuating chromoelectric field. It is revealed that the main contribution here comes 
from the unstable modes which grow exponentially with time.  
A strong direction dependence of the energy loss has also been demonstrated.
\end{abstract}



\end{frontmatter}



One of the goals of the ongoing relativistic heavy ion 
collision experiments at the Relativistic Heavy Ion 
Collider (RHIC) and the experiments at CERN Large Hadron 
Collider (LHC) is to produce quark gluon plasma (QGP) and study its properties.
According to the prediction 
of lattice quantum chromodynamics, QGP is expected to 
be formed when the temperature of nuclear matter is raised above 
its critical value, $T_c\sim 170$ MeV, or equivalently the energy 
density of nuclear matter is raised above $1~GeV/fm^{3}$
~\cite{Karsch02}. The possibility of QGP formation at RHIC 
experiment, with initial density of $5~GeV/fm^{3}$ is supported by the observation of high $p_T$ hadron suppression.
This phenomena commonly known as jet quenching actually is related to
the energy loss of the fast moving partons in the plasma~\cite{jetquen}. Apart form jet-quenching, several possible probes have been studied in order to characterize the 
properties of QGP. 

The estimation of energy loss in quark gluon plasma is thus essential to understand high $p_T$ 
hadron production in relativistic plasma. The first such estimation was performed
by Bjorken \cite{jetquen} long back. Since then, lot of progress has been made to perfect such estimates
under various circumstances including both the collisional and radiative losses 
\cite{collisional}. Most of these
calculations are performed in situations where the distributions of soft partons providing the
thermal background are assumed to have isotropic momentum distribution. In realistic scenario, due to 
rapid initial expansion the system in the longitudinal direction cools faster than the transverse direction 
leading to $\langle p_L^2\rangle \ll \langle p_T^2\rangle $. Such momentum anisotropy might lead to 
collective modes having characteristic behavior distinct from what happens in isotropic 
plasma. This has been vividly exposed in \cite{romatschke03,romatschke04} where it is shown that
these modes might have several branches and some of those might even be unstable. 
The number of unstable modes, as revealed in \cite{romatschke03} depends on whether the system is stretched or 
squeezed along the direction of anisotropicity. The most important among those are the modes which 
grow exponentially in time. These, as we shall see, contribute to the energy loss  when the 
random statistical fluctuations of the fields are considered. 
This particular aspect of energy loss has remained unexplored so far.


In all the calculations except \cite{chakraborty07} the contribution due to field fluctuations have been ignored or eikonal
approximation is assumed. Interestingly, however, in \cite{chakraborty07} 
it has been shown that such a  field fluctuation might
lead to an energy gain of a moving quark in isotropic QGP irrespective of its 
velocity. It would therefore be interesting
to see what happens in situations where the ground state is anisotropic and the random behavior of the chromodynamic fields 
are considered. This is precisely what we consider in this letter. This contribution is time dependent and  
dominates at the early stage of the collision when the anisotropicity exists and appear as corrections to the polarization
loss. The latter, {\em i.e.} the polarization loss however, remain  unaffected by the unstable modes 
for reasons explained in \cite{romatschke05, romatschke052}. 

The calculations of fluctuation spectrum of
chromodynamic fields even for an equilibrium
system is a daunting task because of numerous terms to be
taken into account. For anisotropic system to be realized
in relativistic heavy ion collisions, it becomes quite
difficult to calculate the field correlations. Instead of considering
momentum distribution corresponding to a QGP likely
to be produced in relativistic heavy ion collisions we concentrate
here on a two-stream plasma for the sake of tractability of our
calculations and to bring the underlying dynamics to clearer relief. 
The two stream plasma is unstable with respect to both electric and magnetic interactions.The effect of these
unstable modes in a 
two stream plasma has been studied recently \cite{stanislaw06,stanislaw081,stanislaw082,stanislaw10,majumder09}. In \cite{majumder09} they have shown that the 
transport coefficient grows exponentially in
time due to these unstable modes.
In the present work, we shall be concentrating on the contributions to the
energy change of a fast moving parton due to chromoelectric field
fluctuations which, as indicated before, remained unaddressed so far 
even in the calculations of
quark energy loss in anisotropic medium. In particular, 
we shall show that the 
unstable modes lead to an exponentially growing energy change due
to fluctuations in a two-stream plasma.

When a charge particle moves through a plasma it loses part of its energy 
due to its interaction with medium particles~\cite{akhiezer,sitenko}. The energy loss of a particle
is determined by the breaking forces acting on the particle due to the 
chromoelectric field produced by the particle itself while moving. The energy loss of the 
moving particle per unit time :
\be
\frac{dE}{dt}=Q^a{\bf v.E}^a|_{{\bf r}={\bf v} t}.\label{eq1}
\ee
If we take into account the fluctuation of the field in the plasma and also the change 
of velocity of the particle, that is, the particle recoil in collision,
Eq.(\ref{eq1}) is modified and we get~\cite{akhiezer,sitenko} 
\be
\frac{dE}{dt} =\langle Q^a{\bf v}(t).{\bf E}^a({\bf r}(t),t) \rangle,\label{eq2}
\ee
where $\langle...\rangle$ denotes statistical average. ${\bf E}({\bf r}(t),t)$, 
the chromoelectric field which is a random function of position and time.

The classical equation of motion of the particles in the electromagnetic 
field has the form 
\be
\frac{d{\bf p}}{dt}=Q^a[{\bf E}^a( {\bf r}(t),t)+ {\bf v}\times{\bf  B}^a({\bf  r}(t),t)].\label{eq3}
\ee
Integrating the above equation, we obtain
\bea
{\bf  v}(t)= {\bf v}_0+\frac{1}{E_0}\int_0^t dt_1 Q^a {\bf  F}^a({\bf  r}(t_1),t_1),\nn\\
 {\bf r}(t)= {\bf v}_0t+\frac{1}{E_0}\int_0^t dt_1\int_0^{t_1} dt_2 Q^a {\bf  F}^a( {\bf r}(t_2),t_2).\label{eq4}
\eea
where ${\bf  r}_0$ and $ {\bf v}_0$ are the radius vector and the velocity 
at the initial stage of the particle. $E_0$ is the initial parton energy and 
${\bf  F}= {\bf E}+{\bf  v}\times {\bf  B}$.

We choose a time interval $\Delta t$ sufficiently large compared with 
the period of 
the random fluctuation of the particle field in the plasma but 
small compared to the  time during which the motion of the particle 
changes appreciably. During this time interval $\Delta t$ the particle
trajectory differs little from straight line. So we can write only the 
leading order
terms for the particle velocity and the field acting upon the particle 
at this time interval:
\bea
{\bf  v}(t)&=& {\bf v}_0+\frac{1}{E_0}\int_0^t dt_1 Q^a {\bf  F}^a({\bf r}(t_1),t_1),\nn\\
 {\bf E}^a( {\bf r}(t),t)&=& {\bf E}^a( {\bf r}_0(t),t)+\frac{Q^a}{E_0}\int^t_0dt_1\int ^{t1}_0dt_2\sum_jE^b_{j}({\bf r}_0(t_2),t_2)\frac{\partial}{\partial r_{0j}}{\bf E}^a({\bf r}_0(t),t)\label{eq5}
\eea
where, ${\bf r}_0(t)={\bf v}_0 t$. Substituting Eq.(\ref{eq5}) into Eq. (\ref{eq2}) and using 
$\langle E^a_iB^a_j\rangle_{\beta}=0$~\cite{landau}, we find,
\bea
\frac{dE}{dt}=\langle Q^a{\bf v}_0 . {\bf E}^a({\bf r}_0(t),t)\rangle_{\beta}+\frac{Q^aQ^b}{E_0}\int^t_0dt_1\left\langle {\bf E}^b({\bf r}_0(t_1),t_1). {\bf E}^a( {\bf r}_0(t),t)\right \rangle_{\beta}~~~~~~~~~~\nn\\
+\frac{Q^aQ^b}{E_0}\int^t_0dt_1\int^{t_1}_0dt_2\left\langle  \sum_jE^b_{j}( {\bf r}_0(t_2),t_2)\times\frac{\partial}{\partial r_{0j}} {\bf v}_0.{\bf E}^a({\bf r}_0(t),t)\right\rangle_{\beta}\label{eq6}.\eea

The average value of the fluctuating part of the field vanishes
 so that the quantity $\langle  {\bf E}({\bf r}(t),t)\rangle$ is the 
chromoelectric field produced by the particle itself in the plasma. 
Thus, the first term in Eq.(\ref{eq6}) is the polarization energy loss of the 
moving parton calculated in Ref.~\cite{akhiezer,sitenko,gyulassy}. The second 
part
corresponds to the statistical part of the dynamic friction due to the space-time 
correlation of the fluctuations in the electric 
field~\cite{akhiezer,sitenko}. 
The third part determines 
the average change in energy of the moving particle due to the correlation 
between the fluctuation in the velocity of the particle and the fluctuation in the
electric field in the plasma. The presence of such correlations lead to
additional change in the energy of the moving particle.
The calculation of polarization energy loss in anisotropic plasma
to be realized in relativistic heavy ion collision has been
calculated in Ref.~\cite{romatschke05,romatschke052}. 
As mentioned earlier, in this work, we shall 
consider a two stream plasma which permits us to obtain results in closed
form revealing the physical mechanism into clearer focus. Hence we take, 
\be
f({\bf p})=(2\pi)^3n[\delta^{(3)}({\bf p}-{\bf q})+\delta^{(3)}
({\bf p}+{\bf q})]\label{function}
\ee
where, $n$ is the effective parton density in a single stream.
Such a system is extremely unstable and leads to exponentially
growing collective modes for the plasma excitation.
As mentioned earlier, the first term in Eq.(6) corresponds to
the polarization loss which can be calculated in such a system and
the presence of unstable modes can be taken care of as in 
Ref.~\cite{romatschke05,romatschke052}. 
Our
main concern here is to calculate the contribution
due to statistical part where the unstable modes play important
role.
It is to be noted that in non-equilibrium situation it is extremely 
difficult to obtain the chromodynamics 
field~\cite{stanislaw081,stanislaw082,stanislaw10,majumder09} although
in the present case only the chromoelectric field is involved.
We consider only the longitudinal part of the chromoelectric contribution.
The corresponding modes are obtained by setting
the dielectric function, $\epsilon_L(\omega,{\bf k}) = 0$ giving four
roots, $\pm\omega_{\pm}({\bf k})$, where~\cite{stanislaw082,majumder09}
\be
\epsilon_L(\omega,{\bf k})=\frac{(\omega-\omega_+({\bf k}))
(\omega+\omega_+({\bf k}))(\omega-\omega_-({\bf k}))
(\omega+\omega_-({\bf k}))}{(\omega^2-({\bf k.u})^2)^2}\label{epsilon2},
\ee
and
\be
\omega^2_{\pm}({\bf k})=\frac{1}{{\bf k}^2}\Big[{\bf k}^2
({\bf k.u})^2+\mu^2({\bf k}^2-({\bf k.u})^2)\pm\mu
\sqrt{({\bf k}^2-(\bf k.u)^2)(4{\bf k}^2({\bf k.u})^2+\mu^2
({\bf k}^2-(\bf k.u)^2))}\Big].
\ee
Note that, $0<\omega_+({\bf k})\in R$ for any value of $\bf k$ but 
$\omega_-({\bf k})$ is imaginary for ${\bf k}^2({\bf k.u})^2
<2\mu^2({\bf k}^2-(\bf k.u)^2)$ when it 
represents 
the two-stream electrostatic instability generated due to the 
mechanism analogous to the Landau damping. 
For ${\bf k}^2({\bf k.u})^2\geq2\mu^2({\bf k}^2-(\bf k.u)^2)$, 
the mode is stable, $0<\omega_-({\bf k})\in R$.
Let us now consider the domain of wave vectors obeying
 ${\bf k}^2({\bf k.u})<2\mu^2({\bf k}^2-({\bf k.u})^2)$ 
when $\omega_-({\bf k})$ is imaginary and it represents the 
unstable electrostatic mode.
In this case we write down  $\omega_-({\bf k})$ as $i\gamma_k$ with $0<\gamma_k\in R$.

Considering the correlations between the electric field fluctuations
and between the velocity and field fluctuations
coming from the contributions of the unstable mode (which are fastest growing 
function of $(t_1+t_2)$ and $(t_1-t_2)$) the second and the third
 terms in Eq.(6)
can be combined to obtain~\cite{mandal}
\bea
\frac{dE}{dt}&=&\frac{Q^aQ^b}{E_0}\frac{g^2}{4}\delta^{ab}n\int
\frac{d^3k}{(2\pi)^3}\frac{1}{{\bf k}^2×}
\frac{\Big(\gamma_k^2+({\bf k.u})^2\Big)^3\Big(\gamma_k^2-({\bf k.v}_0)^2\Big)}{\gamma_k
\Big(\gamma_k^2+({\bf k.v}_0)^2\Big)^2}
\frac{\rm{e}^{2t\gamma_k}}{(\omega_+^2-\omega_-^2)^2×}\nn\\&&
\label{simple}
\eea
where, we have only considered the fastest growing modes and dropped the
other terms.

\begin{figure}[h]
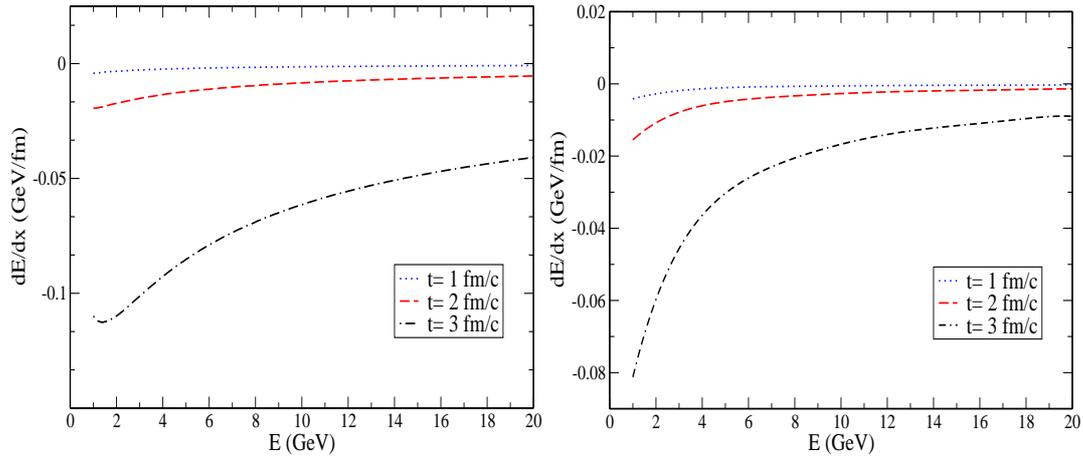

\begin{center}
\epsfig{file=v_parallel_u1.eps,width=7cm,height=6cm,angle=0}
\epsfig{file=v_perp_u1.eps,width=7cm,height=6cm,angle=0}
\end{center}
\caption{(Color online) Energy loss of light partons in a two-stream plasma.
Left (right) panel corresponds to the case when the jet velocity is parallel
(perpendicular) to the anisotropy vector, $u$.}  
\label{fig1}
\end{figure}

As can be seen from Eq.(\ref{simple}) the expression for the change in energy,
in the limit $\gamma_k\,\rightarrow\,0$,
diverges. However, for non-zero $\gamma_k$, as is the case here,
 the effect of the growth
of unstable modes in a two-stream plasma is clearly revealed. 
This contribution to the energy change due to presence of unstable modes
in an expanding plasma should be added to the usual polarization loss.
For numerical estimate we integrate Eq.(\ref{simple}) for light quark. 
Since we are
interested in the relativistic streams, we use $u=1$ for subsequent
calculation.
The anisotropy is determined by a single vector, here the flow velocity
${\bf u}$ which is assumed to be in the $z$-direction. Two cases have
been considered : (i) ${\bf v}_0\parallel{\bf u}$ and
(ii) ${\bf v}_0\perp{\bf u}$. The results are shown in Fig.(\ref{fig1})
for a typical plasma temperature, $T \sim 400$ MeV. Here, we find that the light parton loses energy in a two-stream plasma
due to chromoelectric field fluctuations unlike what has been
observed in case of heavy quark in isotropic plasma~\cite{chakraborty07}.   
It is also observed that the energy loss grows with time.
We also show that the energy loss strongly depends on the
direction of propagation with respect to the anisotropy vector, $u$ 
in this case as reflected in Fig.(\ref{fig1}) and in Eq.(10).
For a parton, the energy loss for a given time is more in the
anisotropy direction than in the perpendicular direction. 
Similar behaviour is observed in Refs.~\cite{romatschke05,romatschke052} while
calculating the polarization loss in anisotropic plasma.


In the present work we estimate the parton energy loss in
two-stream plasma due to chromoelectric field fluctuations. 
In particular, we expose the role of the unstable
modes those are present in such a scenario which changes the results
both qualitatively and quantitatively. In fact, we show that
these modes lead to an exponential growth of the energy change in
presence of the fluctuating fields. For simplicity, only the
longitudinal chromoelectric fields have been considered here. It is found that 
light partons lose energy 
in a two-stream plasma, which further, depends on the
direction of propagation with respect to the anisotropy
vector (direction of the stream velocity in the present case). 
Note that the fields corresponding to the QGP produced
in heavy ion collisions are difficult to obtain with a realistic
anisotropic momentum distribution. However, it will be interesting to 
extend the present calculation in the context
of relativistic heavy ion collisions with realistic anisotropic
momentum distributions of the bath particles. 
Inclusion of such corrections, to the usual polarization energy loss, 
might prove to be an important mechanism for the parton to lose
energy into the plasma, at least initially,
when large anisotropy might exists.

\end{document}